\documentclass[twocolumn,amsmath,amssymb,pra,showpacs]{revtex4}
\usepackage{setspace}
\usepackage{graphicx}
\usepackage{amsmath}
\usepackage{amssymb}
\usepackage{latexsym}

\usepackage{natbib}
\begin{document}

\title{Quantifying Finite Temperature Effects in Atom Chip Interferometry\\of Bose-Einstein Condensates}
\author{R.G. Scott$^{1}$, D.A.W. Hutchinson$^{2}$, T.E. Judd$^{1}$, T.M. Fromhold$^{1}$}
\affiliation{$^{1}$Midlands Ultracold Atom Research Centre, School of Physics and Astronomy, University of Nottingham, Nottingham, NG7 2RD, United Kingdom. \\$^{2}$The Jack Dodd Centre for Quantum Technology, Department of Physics, University of Otago, P.O. Box 56, Dunedin, New Zealand.}
\date{5/01/09}



\begin{abstract}
We quantify the effect of phase fluctuations on atom chip interferometry of Bose-Einstein condensates. At very low temperatures, we observe small phase fluctuations, created by mean-field depletion, and a resonant production of vortices when the two clouds are initially in anti-phase. At higher temperatures, we show that the thermal occupation of Bogoliubov modes makes vortex production vary smoothly with the initial relative phase difference between the two atom clouds. We also propose a technique to observe vortex formation directly by creating a weak link between the two clouds. The position and direction of circulation of the vortices is subsequently revealed by kinks in the interference fringes produced when the two clouds expand into one another. This procedure may be exploited for precise force measurement or motion detection.
\end{abstract}

\maketitle

\section{Introduction}

Recent experiments~\cite{kettnew} have demonstrated the ability to read out the initial relative phase $\Delta$ of two Bose-Einstein condensates (BECs) that are subsequently merged on an atom chip~\cite{hinds,folmanfull,reichelrev,fortaghrev,kruger}. In these experiments, $\Delta$ is inferred by measuring its effect on the size of the cloud after free expansion. Theoretical work~\cite{scottjuddinterf} explained these findings in terms of the resonant production of a black soliton (or $\pi$-state) when $\Delta=\pi$~\cite{scottjuddinterf,negretti,nikuni}. The soliton is unstable, and hence decays via the excitation of stable vortices~\cite{scottol,scottari,scottfull}. Analysis of this mechanism highlighted the potential for exploiting the resonance to develop motion detectors~\cite{scottjuddinterf}, or sensors that could be used to measure weak forces due to gravity, electric, or magnetic fields. The analysis also showed that at finite temperatures, the resonance becomes sufficiently broadened that the number of vortices produced by the merger varies smoothly with $\Delta$, so that all values of $\Delta$ can, in principle, be detected. However, the calculations were unable to ascribe specific temperature values to the different behavioral regimes of the interferometer, nor did they account for the zero-temperature mean-field depletion. Consequently, they did not reveal the limits of the interferometer's sensitivity, nor the temperatures required to achieve precision measurement. Moreover, the proposal for motion detection lacked a mechanism to detect the underlying production of vortices. Further theoretical work is now required to explore the effect of finite temperature on atom-chip interferometry and identify an appropriate read-out protocol for vortex production and motion detection.

The atom clouds in chip traps are typically highly elongated, and can therefore be used to explore a one-dimensional regime, in which $k_{B}T$ and $\mu$ are both less than $\hbar \omega_{r}$, where $T$ is the temperature of the cloud, $\mu$ is its chemical potential, and $\omega_{r}$ is the radial (high) trapping frequency~\cite{Hofferberth,Hofferberth2}. For the parameters considered in this paper, and in previous related experiments~\cite{kettnew}, this condition is \emph{not} satisfied because $\mu> \hbar \omega_{r}$. This means that radial excitation of the cloud, namely soliton and vortex production, can occur. By varying $T$, we may access different regimes of phase coherence. Global phase coherence in elongated clouds only exists below a characteristic temperature $T_{\phi}$~\cite{petrov,aspect}, which may be much less than the BEC transition temperature $T_{c}$. At temperatures below $T_{c}$, but \emph{not} below $T_{\phi}$~\cite{petrov,aspect}, there exists a ``quasicondensate'' regime, where the phase coherence length is less than the axial length of the atom cloud. The temperature $T_{\phi}$ is zero for uniform one-dimensional Bose gases, and increases as the axial length decreases. For interferometry purposes, we require that $T<T_{\phi}$, so that there is global phase coherence and, as a result, $\Delta$ is well-defined. However, even at such temperatures, which are the focus of this paper, we find that low-energy thermal excitations along the axial direction still play a crucial role in the function of the interferometer.


In this paper, we present detailed calculations, which now enable us accurately to characterize the behavior of the interferometer as a function of $T$. This is done by developing a finite temperature truncated Wigner method~\cite{Janne,Janne2}, which includes thermal fluctuations through the excitation of Bogoliubov modes. When $T<<T_{\phi}$, the phase fluctuations are set by the zero-temperature mean-field depletion of the condensate. We show that these fluctuations are small ($\sim 0.03 \pi$) for the parameters used in the experiment~\cite{kettnew}, and hence the resonant production of vortices is sharply peaked around $\Delta \simeq \pi \pm 0.05 \pi$. Consequently, this type of atom interferometer could be exploited to precisely detect when $\Delta=\pi$, which is potentially useful for the measurement of motion, small forces, and spatially-varying phase profiles. We show that for larger $T$, comparable to, yet still below, $T_{\phi}$, global phase coherence is preserved, but axial thermal phase fluctuations are significant. This causes vortex production to vary more smoothly with $\Delta$, thus creating a sensor capable of detecting all values of $\Delta$. For the parameters used in the experiment~\cite{kettnew}, this behavioral regime occurs when $T_{\phi}>T \gtrsim 90$ nK. 

We also propose a technique to detect vortex production directly by performing a partial merge of the atom clouds. This method works by establishing a weak link between the two BECs, which is a topic of broad interest due to its implications for Josephson oscillations, the Sine-Gordon equation, and questions of coherence~\cite{bouchoule,polkovnikov,Hofferberth,whitlock}. The partial merge allows vortices to form in the region of low density between the two BECs. Presence of vortices is then detected after expansion by the appearance of abrupt kinks in the interference fringes, which not only map out vortex production along the cloud, but also provide a mechanism to read out the relative velocity of the two BECs, acting as a motion detector. 

The paper is organized as follows. In section~\ref{system} we describe the system and explain our theoretical model. In section~\ref{resultA} we explore the operation of the interferometer as a function of $T$. In section~\ref{resultB} we discuss the partial merge technique and possible applications in motion sensing. In section~\ref{conc} we summarize our findings and conclude.

\section{System and methodology}
\label{system}

In the experiments~\cite{kettnew}, a $^{23}$Na BEC containing $2N_{T} = 4\times 10^{5}$ atoms is prepared in an atom chip trap, then split equally into two clouds. We assume that the splitting process is carried out adiabatically, and hence construct two finite $T$ initial states~\cite{foot4}, each containing $N_{T}$ atoms, for every simulation. We believe this to be a reasonable assumption because the split was carried out over 75 ms in the experiment~\cite{kettnew}. 

 
The trap frequencies in the axial and radial directions are $\omega_{z} = 2\pi \times 9$ rad s$^{-1}$ and $\omega_{r} = \omega_{x} = \omega_{y} = 2\pi \times 1000$ rad s$^{-1}$ respectively, creating two clouds of peak density $2.2 \times 10^{20}$ m$^{-3}$ at $T=0$ in the double-well potential. For these parameters, $T_{\phi} = 15N_{T}\left(\hbar\omega_{z}\right)^{2}/32\mu k_{B} \approx 100$ nK~\cite{petrov,aspect}, where $\mu = 3500$ Hz is the chemical potential.  

Many models of cold atom clouds have been developed, and each have strengths and weaknesses, meaning that they perform well under different conditions~\cite{BlairRev,NickRev}. For example, the traditional zero $T$ truncated Wigner method~\cite{steel,sinatra}, because of its inclusion of quantum noise, has been very successful in describing spontaneous processes like the formation of scattering halos~\cite{meotago}, or the suppression of Cherenkov radiation~\cite{meCheren}. However, in this study, we must simulate the thermal excitations of elongated atom clouds, which are long-wavelength axial sound-like modes, requiring a finite $T$ approach. Firstly, we note that this atom chip system is well-suited to classical-field methods~\cite{BlairRev,NickRev}, which include low-energy modes that have an occupation much larger than one particle. Since the modes are highly Bose-degenerate, the matter-wave field behaves much like a classical field. It is therefore valid to model the condensate plus the low-energy excited modes within a Gross-Pitaevskii formalism, which propagates a classical field, and, consequently, can describe all classical aspects of a finite $T$ BEC. Secondly, a stochastic approach is appropriate to capture the thermal fluctuations. As a result of these two requirements, three possible methods are: the projected Gross-Pitaevskii equation~\cite{blakie,Davisnat}, the stochastic Gross-Pitaevskii equation~\cite{stoof}, or the finite $T$ truncated Wigner method~\cite{Janne,Janne2}. All three of these methods are, in principle, applicable. However, they differ in that the projected Gross-Pitaevskii equation and the stochastic Gross-Pitaevskii equation treat all fluctuations thermally in equilibrium, whilst thermal \emph{and} quantum fluctuations are added explicitly in the finite $T$ truncated Wigner method. This means that the finite $T$ truncated Wigner method describes the atom cloud well at the low temperatures suitable for performing interferometry. In addition, it can predict the limits of the interferometer sensitivity at $T=0$ (discussed in section~\ref{zeroTsec}). For these reasons, we use the finite $T$ truncated Wigner method in this study.

In this paper, we consider only the axial excitations of the BEC, and neglect radial excitations. This approximation is valid if the mean thermal occupation of the first excited state in the radial direction $\bar{N}^r_1<1$, where
\begin{equation}
\bar{N}^r_1 = \frac{1}{e^{E^r_1/k_B T}-1},
\end{equation}
in which $E^r_1 \approx 1.5\hbar\omega_r$ is the energy of the first excited radial mode. This condition is satisfied provided
\begin{equation}
T<E^r_1/k_B\ln\left(2\right) \approx 100 \; \mbox{nK}.
\label{condition}
\end{equation}
In this paper, we explore the temperature range $T<T_{\phi}\approx 100$ nK, where inequality~\ref{condition} is satisfied.

\begin{figure}[tbp]
\centering
\includegraphics[width=0.7\columnwidth]{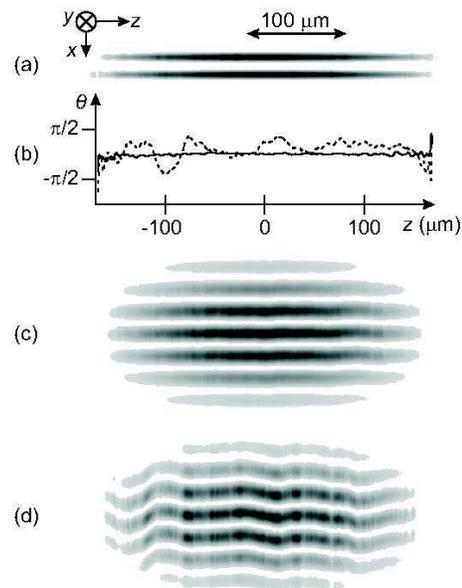}
\caption{(a) Initial ($t=0$) atom density profile of the two BECs in the $y=0$ plane (axes inset) at $T=90$ nK. (b) Typical initial phase profiles $\theta\left(z\right)$ along one of the BEC's ($z$) axis of cylindrical symmetry at $T=0$ (solid curve) and $T=90$ nK (dashed curve). (c) Interference fringes in the $y=0$ plane [axes inset in (a)] after release from the double-well potential and 8 ms free expansion (Process A) at $T=0$. (d) As (c), but for $T=90$ nK. Horizontal bar in (a) shows scale of (a), (c) and (d).}
\label{f1}
\end{figure}

In order to construct the finite $T$ state, we first solve the three-dimensional Gross-Pitaevskii equation to obtain the initial density profile $n_{0}\left(x,y,z\right)$ of the condensate mode $\Psi_{0}$, using the coordinates defined in Fig.~\ref{f1}(a). The low-energy axial excitations are then calculated by solving the one-dimensional Bogoliubov-de Gennes equations
\begin{equation}
\left[H_{\mbox{sp}}+2U_{0}n_{0}\left(0,0,z\right)-\mu\right]u_{j} - U_{0}n_{0}\left(0,0,z\right)v_{j} = E^{z}_{j}u_{j}
\label{eq:Bog1}
\end{equation}
and 
\begin{equation}
-\left[H_{\mbox{sp}}+2U_{0}n_{0}\left(0,0,z\right)-\mu\right]v_{j} + U_{0}n_{0}\left(0,0,z\right)u_{j} = E^{z}_{j}v_{j} ,
\label{eq:Bog2}
\end{equation}
where 
\begin{equation}
H_{\mbox{sp}} = -\frac{\hbar^{2}}{2m}\frac{d^{2}}{dz^{2}} + \frac{1}{2}m \omega_{z} z^{2}
\end{equation}
and $U_{0} = 4\pi\hbar^{2}a/m$, in which $E^{z}_{j}$ is the energy of the $j^{\mbox{th}}$ Bogoliubov mode, $m$ is the mass of a single $^{23}$Na atom, and $a=2.9$ nm is the s-wave scattering length. Since Eqs.~\ref{eq:Bog1} and \ref{eq:Bog2} are one-dimensional, the eigenvalue $E_j^z$ does not include the contribution from the radial kinetic energy, which we calculate as $1.64 \times 10^{-31}$ J from the Gross-Pitaevskii equation. Consequently the true eigenvalue of the $j^{\mbox{th}}$ Bogoliubov mode is $E_j = E_j^z + 1.64 \times 10^{-31}$.

We find that the low-energy Bogoliubov modes are long-wavelength, phonon-like excitations. This implies that the use of plane-wave noise to model a finite $T$ BEC in a previous study~\cite{scottjuddinterf} was a reasonable approximate method. However, the use of plane-wave noise meant that, in contrast to the present work, the temperature of the BEC could not be specified.

The $j^{\mbox{th}}$ Bogoliubov mode is defined as 
\begin{equation}
\psi_j = A\left(x,y\right)\left[u_j\left(z\right)\beta_j - v_j^\ast\left(z\right)\beta_j^\ast\right] ,
\end{equation}
where 
\begin{equation}
A\left(x,y\right) = \sqrt{\frac{n_{0}\left(x,y,0\right)}{\int n_{0}\left(x,y,0\right) dx dy}},
\end{equation}
and $\beta_{j}$ is a random complex number such that
\begin{equation}
\overline{\beta_j^\ast \beta_j}=\bar{N}_j+1/2 ,
\label{eq:distribution}
\end{equation}
in which 
\begin{equation}
\bar{N}_j = \frac{1}{e^{E_j/k_B T}-1} 
\end{equation}
is the mean thermal occupation of the $j^{\mbox{th}}$ mode. The $1/2$ in Eq.~\ref{eq:distribution} describes the zero $T$ mean-field depletion of the BEC, and the quantum vacuum fluctuations~\cite{meotago}. The finite $T$ initial state is then constructed as 
\begin{equation}
\Psi = \Psi_0 + \sum_{j} \psi_j .
\label{eq:initial}
\end{equation}
This is a finite-temperature truncated Wigner method~\cite{Janne,Janne2}.

We must choose a $j$ value at which we truncate the summation in Eq.~\ref{eq:initial}. As $E_j$ increases, $\bar{N}_j$ decreases, meaning that $\overline{\beta_j^\ast \beta_j}$ and, consequently, the average excitation occupation $\overline{\int\psi_j^{\ast}\psi_j dz}$ also decreases. In this paper we neglect quantum vacuum fluctuations~\cite{meotago}, since a previous study has shown them to have no effect in this system~\cite{scottjuddinterf}. Consequently, we stop adding noise once the average excitation occupation $\overline{\int\psi_j^{\ast}\psi_j dz}$ drops below one atom.

In our simulations, at time $t=0$ the two finite $T$ initial states are at rest in a double-well potential~\cite{foot4}, which has minima at $x=\pm 3$ $\mu$m [Fig.~\ref{f1}(a)]. In Fig.~\ref{f1}(b), we characterize the initial phase fluctuations by plotting typical phase profiles $\theta\left(z\right)$ along one BEC's axis of cylindrical symmetry, for $T=0$ (solid curve) and $T=90$ nK (dashed curve). The phase fluctuations at $T=0$ are due to mean-field depletion of the condensate, and are very small in amplitude ($\sim 0.03 \pi$), except at the very edges of the BEC where the atom density is low. The phase fluctuations in the BEC at 90 nK are much larger, typically $\sim 0.3 \pi$ in the centre of the BEC, but approaching $\pi$ at the extremities. Phase fluctuations of amplitude $0.5 \pi$ in each BEC implies a potential \emph{relative} phase fluctuation of $\pi$ if the local fluctuations in each BEC are of opposite sign, sufficient to switch the system to and from resonance (vortex production at $\Delta=\pi$) and anti-resonance ($\Delta = 0$).  

In this paper, we consider three distinct recombination procedures, which generate different dynamics. In this section, we focus on the first of these procedures, henceforth called Process A. We introduce Processes B and C in Sections~\ref{resultA} and ~\ref{resultB} respectively.

In Process A, we abruptly release the two BECs from the double-well potential at $t=0$, and allow them to expand into one other for 8 ms. The resulting interference patterns enable us to relate the phase fluctuations to an experimentally observable effect, specifically the modulation of the fringe positions~\cite{kettnew2}. Figure \ref{f1}(c) shows interference fringes produced, for $T=0$, by Process A. The fringes are almost straight and at all points parallel to the axial ($z$) direction, indicating that the relative phase is approximately constant and the phase fluctuations are small ($<<\pi$). In contrast, at $T=90$ nK [Fig.~\ref{f1}(d)] the position of the density nodes and anti-nodes varies along $z$, so that the fringes have an undulating appearance, particularly at the extremities of the cloud where the density is low. The interference pattern in Fig.~\ref{f1}(d) is similar to that shown in experiment~\cite{kettnew2}, with comparable variations in fringe positions, suggesting that the temperature of the BEC used in the experiments is $\sim 90$ nK.



\section{Production of vortices}
\label{resultA}

In this section, as in experiment~\cite{kettnew}, we merge the two BECs by smoothly transforming the double-well potential into a single-well potential. This merging procedure is referred to as Process B. The operation is performed over 5 ms, since experiments~\cite{kettnew} and our previous theoretical studies~\cite{scottjuddinterf} have shown this to be a suitable timescale to observe vortex formation. However, the merging time is not crucial as similar dynamics occur over a wide range of merging times from $\sim 2$ to $\sim 20$ ms. Different behavioral regimes may accessed by selecting merging times outside these limits~\cite{scottjuddinterf}. We now analyze how the dynamics change with temperature.

\subsection{Zero temperature dynamics}
\label{zeroTsec}

\begin{figure}[tbp]
\centering
\includegraphics[width=0.7\columnwidth]{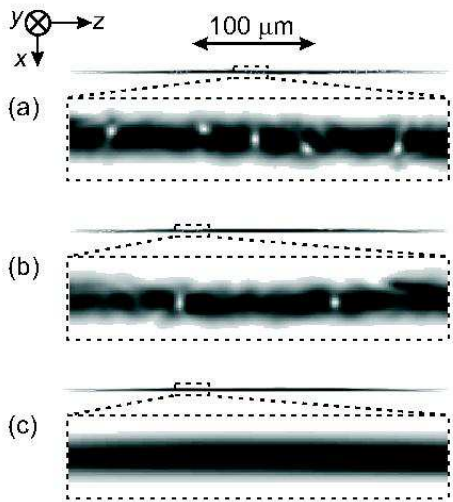}
\caption{Atom density profiles in the $y=0$ plane (axes inset) calculated for $T=0$ at $t=5$ ms when the double-well potential has been smoothly transformed into a single-well potential (Process B) for $\Delta = \pi$ (a), $0.95 \pi$ (b) and $0.9 \pi$ (c). Density profiles within the regions enclosed by the small dashed rectangles are shown enlarged. Horizontal bar shows scale.}
\label{f2a}
\end{figure}

Firstly, we reproduce the resonant production of vortices at $T=0$ reported in Ref.~\cite{scottjuddinterf}, by smoothly merging the two BECs via Process B. Figure \ref{f2a}(a) shows a typical merged cloud at $t=5$ ms, for $\Delta=\pi$. Vortices have been created along the entire length of the cloud. Five of these can be seen as low density (white) spots within the central region shown enlarged within the lower dashed box. When $\Delta$ is reduced to $0.95 \pi$, vortices are still present, but they are now concentrated towards the extremities of the cloud, as shown Fig.~\ref{f2a}(b). This behavior occurs because the phase fluctuations are larger at the extremities of the cloud [see Fig.~\ref{f1}(b)] and, consequently, the $\pi$-state may be formed locally, if the phase fluctuations in each BEC combine to create a net \emph{relative} phase fluctuation exceeding $0.05 \pi$. However, when we reduce $\Delta$ to $0.9 \pi$ [Fig.~\ref{f2a}(c)], no vortices are produced at any point in the cloud. This is because, as shown in Fig.~\ref{f1}(b) (solid curve), the phase fluctuations are of order $0.03 \pi$ in this elongated geometry at $T=0$. Hence we determine the limits of the sensitivity of this interferometry mechanism: for the parameters used in the experiments~\cite{kettnew}, the resonant production of vortices at $T=0$ requires that $\Delta \simeq \pi \pm 0.05 \pi$.

\begin{figure}[tbp]
\centering
\includegraphics[width=0.6\columnwidth]{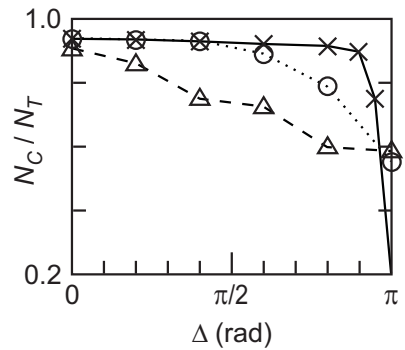}
\caption{$N_{C}/N_{T}$ versus $\Delta$ curves at $t=5$ ms when the double-well potential has been smoothly transformed into a single-well potential (Process B) for $T=0$ (solid curve with crosses) and $T=60$ nK (dotted curve with circles) and $T=90$ nK (dashed curve with triangles).}
\label{f3}
\end{figure}

As shown in our previous study~\cite{scottjuddinterf}, the vortices are described by high momentum plane-wave modes, so their production can be detected by observing a broadening in momentum space. This increased width in momentum space drives an enhanced spreading of the merged cloud when released from trapping potential, as detected experimentally~\cite{kettnew}. We quantify this effect by integrating the atom density of the merged cloud along the $x$ and $y$ directions [see Fig.~\ref{f1} for axes], then calculating the one-dimensional Fourier power spectrum, $F\left(k_{z}\right)$. We then define the quantity $N_{c} = \int^{k_{c}}_{-k_{c}}F\left(k_{z}\right)dk_{z}$, which is the number of atoms in the merged cloud with $\left|k_{z}\right|<k_{c} = 7 \times 10^{4}$ m$^{-1}$~\cite{scottjuddinterf}, or, equivalently, the number of atoms in the merged cloud moving at speeds less than 0.2 mm s$^{-1}$ in the $z$ direction~\cite{foot2}. The solid curve in Fig.~\ref{f3} shows $N_{c}/N_{T}$ plotted as a function of $\Delta$. When $\Delta=0$, no vortices are produced in the merged cloud, so the BEC has a low internal kinetic energy, and, consequently, $N_{c}/N_{T} \approx 1$. However, when $\Delta=\pi$, the merged cloud contains many vortices. As a result, the BEC has a much larger internal kinetic energy, and hence $N_{c}/N_{T}<<1$. The dramatic drop in $N_{c}$ when $\Delta$ approaches $\pi$ indicates the sharp resonant excitation of the cloud due to the formation of vortices. 

\subsection{Dynamics at finite temperature}

\begin{figure}[tbp]
\centering
\includegraphics[width=0.7\columnwidth]{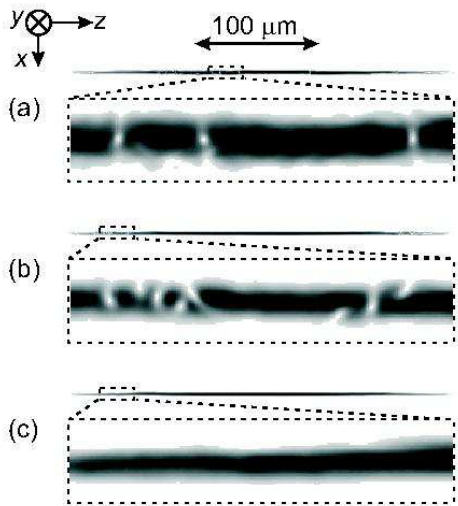}
\caption{Atom density profiles in the $y=0$ plane (axes inset) calculated for $T=60$ nK at $t=5$ ms when the double-well potential has been smoothly transformed into a single-well potential (Process B) for $\Delta = \pi$ (a), $0.6 \pi$ (b) and $0.4 \pi$ (c). Density profiles within the regions enclosed by the small dashed rectangles are shown enlarged. Horizontal bar shows scale.}
\label{f2}
\end{figure}

We now increase $T$ to 60 nK. Again we find vortex production for $\Delta = \pi$, as shown in Fig.~\ref{f2}(a). Vortices form along the entire length of the cloud, and three of these are shown in the enlargement of the region contained within the small dashed rectangle near the center of the cloud. However, as $\Delta$ is reduced the number of vortices in the merged cloud decreases much more gradually than for $T=0$. Even when $\Delta=0.6 \pi$, vortex formation still persists near the edges of the BEC where the fluctuations are largest [Fig.~\ref{f2}(b)]. Vortex formation is finally suppressed when $\Delta$ is reduced to $0.4 \pi$ [Fig.~\ref{f2}(c)]. 

We quantify this smoothing of the resonance in Fig.~\ref{f3}. Since the thermal fluctuations are random, and change from one simulation to another, the number of vortices produced at each value of $\Delta$ also changes between simulations. Therefore, we perform the merger five times at each value of $\Delta$ with different initial thermal fluctuations, then calculate the mean value of $N_{c}$, in order to obtain an averaged result. The dotted curve in Fig.~\ref{f3} shows how the averaged $N_{c}$ varies with $\Delta$ at $T=60$ nK. The curve reveals that the resonance has been significantly broadened compared to the $T=0$ curve (solid), meaning that, when $T=60$ nK, there is significant excitation of the cloud even for $\Delta$ as small as $0.6 \pi$. We also note that the range of $N_{c}$ has been reduced, indicating that this effect will become more difficult to detect as $T$ increases. 

Finally, we increase $T$ to 90 nK. At this temperature we find that $N_{c}$ varies smoothly with $\Delta$ (see the dashed curve in Fig.~\ref{f3}), and we note that the variation in $N_{c}$ is again reduced. As $T$ approaches $T_{\phi} \approx 100$ nK, this variation continues to decrease, until $N_{c}$ is independent of $\Delta$ at $T_{\phi}$. These results are consistent with the findings of a recent experiment~\cite{kettnew}, which reported a gradual increase in the size of the cloud, after expansion, as $\Delta$ approached $\pi$. The calculations of $N_{c}$ presented in Fig.~\ref{f3} suggest that the BEC in the experiment was at $\sim 90$ nK.

\section{Direct detection of vortices}
\label{resultB}

Previous experiments detected the excitation of the BEC by measuring the size of the merged cloud after a free expansion. Excitation was inferred by counting atoms beyond the expanded Thomas-Fermi radius. In order to develop BEC-based atom interferometers as working sensors (see, for example, Refs.~\cite{kettnew,scottjuddinterf,schmiedreview,oberthaler}), it would be desirable to have a more direct mechanism for detecting vortex production. This is essential for the development of such atom interferometers for motion detection, as suggested in a previous study~\cite{scottjuddinterf}.

We find that direct detection of vortex formation can be achieved by performing a partial merge, which we refer to as Process C, such that a weak link is created between the two clouds. To achieve this, in our calculations the full merge from the double-well to single-well potential is halted at $t=2.1$ ms, when the two wells are separated by 3.5 $\mu$m. The BEC is then held for a further 1.9 ms. Under these conditions, the black soliton may still form, and produce vortices in the low density region between the density peaks located at the two minima of the potential~\cite{gunn}. The advantage of the partial merge is that the two peaks in density may still create interference fringes, albeit with a reduced contrast, as they expand rapidly into one another, 8 ms after release from the trapping potential. The presence of vortices will then be revealed as abrupt kinks in the interference pattern, which have been used to detect the formation of vortices in quasi-two-dimensional BECs~\cite{dalibard}. 

\begin{figure}[tbp]
\centering
\includegraphics[width=0.9\columnwidth]{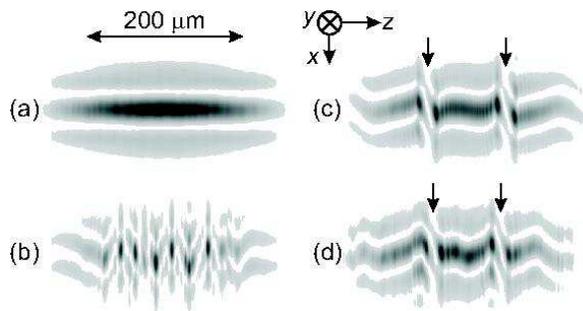}
\caption{Density profiles of BECs in the $y=0$ plane (axes inset) after partial merge and 8 ms free expansion (Process C). (a) $T=0$ and $\Delta=0$. (b) $T=0$ and $\Delta=\pi$. (c) $T=0$ and $\Delta=6 \times 10^{4} z$. (d) As (c), but for $T=90$ nK. Horizontal bar shows scale. Arrows in (c) and (d) indicate positions of two kinks in the fringes caused by vortices.}
\label{f4}
\end{figure}

We initially study this procedure for $\Delta=0$, at $T=0$. The resulting expanded cloud is shown in Fig.~\ref{f4}(a). As expected, the two BECs combine to form a cloud with a smooth density profile, and no vortices are formed during the merger. Consequently, the expanded cloud is a simple low-contrast interference pattern. 

We now repeat the calculation for $\Delta=\pi$. In this case, many vortices form between the two merging BECs. As a result, when the cloud is released from the potential, the interference pattern contains many abrupt kinks [shown in Fig.~\ref{f4}(b)], indicating the position of each vortex. At $T=0$, so many vortices are produced that the expanded cloud is barely recognisable as an interference pattern. We note that the disruption is limited to the central section of the interference pattern. This is because the weak link is \emph{not} formed at the extremities of the cloud, where the two initial BECs have a narrower radial width. 

This procedure provides a mechanism to detect relative motion of the BECs, possibly in response to a weak force. We consider an example in which the BEC at larger $x$ has an initial velocity $v = 0.2$ mm s$^{-1}$ in the $z$ direction. We set the initial phase of the second BEC, at smaller $x$, to be uniformly zero, meaning that $\Delta = mvz/\hbar = 6 \times 10^{4} z$. When we perform the partial merge and expansion, we observe two kinks in the final interference pattern [arrowed in Fig.~\ref{f4}(c)], at positions $z = \pm 52$ $\mu$m, where $\Delta = \pm \pi$. By measuring the distance, $\delta z$, between these kinks, the initial relative speed of the BECs follows directly from $v = h/m \delta z$. Furthermore, the orientation of the kink, namely whether it is a displacement in the positive or negative $x$ direction, indicates the rotation of the vortex, which in turn indicates the direction of $v$ before merging. In Fig.~\ref{f4}(c), both kinks are displacements in the positive $x$ direction, indicating abrupt rotations of $\Delta$ caused by vortices circulating in an anticlockwise direction. As shown in a previous study~\cite{scottjuddinterf}, anticlockwise rotation in the $y=0$ plane infers positive $\partial\Delta/\partial z$, and hence a positive $v$.

This technique is robust to the inclusion of thermal phase fluctuations. To illustrate this, we repeat the above calculation at $T=90$ nK, and present a typical result in Fig.~\ref{f4}(d). Again we see two kinks in the interference pattern at $z = \pm 52$ $\mu$m, although there are some small variations in fringe position compared to the $T=0$ result in Fig.~\ref{f4}(c), particularly at the extremities of the cloud where the phase fluctuations are most significant. At temperatures $T_{\phi}>T>90$ nK, larger phase fluctuations cause several vortices to be created in extended regions around the two positions $z = \pm 52$ $\mu$m. Motion detection is still possible in this regime, but averaging over many absorption images may be necessary to obtain accurate results.

\section{Conclusions}
\label{conc}
Atom chips are a versatile tool which provide the fine control of the splitting and merging of atom clouds required for matter-wave interferometry~\cite{kettnew,andreas1full,andreasdiff2,scottjudddiff}. However, the very elongated geometries that they typically produce have both advantages and disadvantages. A large axial ($z$) length is desirable because it provides natural averaging over any small fluctuations in the trap potentials, the force to be measured, or thermal fluctuations within the BEC. Consequently, the sensitivity can, in some interferometric schemes, be increased by integrating the atom density along the axial direction. Furthermore, elongated clouds are, in principle, sensitive to very weak forces, which create small phase gradients. 

In this paper, we have explored the effect of the elongated geometry on chip-based interferometry by making realistic calculations of the excited states, thus enabling the effect of finite $T$ on the action of the interferometer to be quantified. We have shown that finite $T$ effects are particularly important for elongated BECs due to the low energy of the axial modes. Consequently, BECs in chip traps allow the initial relative phase of the atom clouds to be measured over a wide range, owing to the thermal broadening of the resonant production of vortices. It remains a challenge for experimentalists to achieve temperatures where the resonance is sharp, and exploit it, either for precise force measurement or motion detection. This regime may be accessed by reducing the aspect ratio, but at the expense of losing the advantages of elongated clouds.

\begin{acknowledgments}
This work is funded by EPSRC, the AFOSR through project AOARD-094016, and by the Government of New Zealand through the Foundation for Research, Science and Technology under New Economy Research Fund contract NERF-UOOX0703. We thank J. Ruostekoski and A.D. Martin for helpful discussions. 
\end{acknowledgments}

\bibliography{biblio}

\begin{thebibliography}{40}
\bibitem[{foo({\natexlab{a}})}]{foot4}
\bibinfo{note}{It should be noted that in the experiment temperature was
  measured before splitting, whilst we define temperature after splitting.
  There may be significant cooling due to the splitting process, during which
  the volume of the BEC approximately doubles, and hence the energy
  approximately halves.}
\bibitem[{foo({\natexlab{b}})}]{foot2}
\bibinfo{note}{As the vortices are approximately cylindrically symmetrical
  along the $y$-direction, their associated Fourier components are almost
  identical in the $x$- and $z$-directions.}
\end{thebibliography}

\end{document}